\documentclass[fleqn,10pt]{wlscirepav}
\usepackage[utf8]{inputenc}
\usepackage[T1]{fontenc}

\usepackage{acronym}
\usepackage{amsmath}
\usepackage{amssymb}
\usepackage{color}
\usepackage{hyperref}
\usepackage{graphicx}

\usepackage[normalem]{ulem}

\title{Ultra-low frequency gravitational waves from cosmological and astrophysical processes}

\author{Christopher J. Moore}
\author{Alberto Vecchio}

\newcommand{\bham}{Institute for Gravitational Wave
Astronomy \& School of Physics and Astronomy, 
University of Birmingham, Edgbaston, Birmingham 
B15 2TT, UK}

\affil{\bham}

\begin{document}

\flushbottom
\maketitle
%
%
\vspace{-1cm}
\noindent
\textbf{
    Gravitational waves (GWs) at ultra-low frequencies (${\lesssim 100\,\mathrm{nHz}}$) are key to understanding the assembly and evolution of astrophysical black hole (BH) binaries with masses $\sim 10^{6}-10^{9}\,M_\odot$ at low redshifts\cite{2008MNRAS.390..192S, 2013MNRAS.433L...1S, 2017MNRAS.471.4508K}. 
    These GWs also offer a unique window into a wide variety of cosmological processes\cite{
    PhysRevLett.126.041303, 
    PhysRevLett.126.051303, 
    PhysRevLett.126.041304, 
    PhysRevLett.126.041305, 
    2021PhLB..81636238N,    
    2021ScPP...10...47R,    
    2021SCPMA..6490411A,    
    PhysRevLett.115.181101}.
    Pulsar timing arrays\cite{1978SvA....22...36S, 1979ApJ...234.1100D, 1990ApJ...361..300F} (PTAs) are beginning to measure\cite{2020ApJ...905L..34A} this stochastic signal at $\sim 1-100\,\mathrm{nHz}$ and the combination of data from several arrays\cite{2015Sci...349.1522S, 2015MNRAS.453.2576L, 2018ApJ...859...47A, 2016MNRAS.458.1267V} is expected to confirm a detection in the next few years\cite{2021ApJ...911L..34P}.
    The dominant physical processes generating gravitational radiation at $\mathrm{nHz}$ frequencies are still uncertain.
    PTA observations alone are currently unable\cite{NANOGrav_PT} to distinguish a binary BH astrophysical foreground\cite{2021MNRAS.502L..99M} from a cosmological background due to, say, a first order phase transition at a temperature $\sim 1-100\,\mathrm{MeV}$ in a weakly-interacting dark sector\cite{2021PhLB..81636238N, 2021ScPP...10...47R, 2021SCPMA..6490411A, PhysRevLett.115.181101}.
    This letter explores the extent to which incorporating integrated bounds on the ultra-low frequency GW spectrum from any combination of cosmic microwave background\cite{2006PhRvL..97b1301S, 2020JCAP...10..002C}, big bang nucleosynethesis\cite{2014ApJ...781...31C, Maggiore:2018sht} or astrometric\cite{2018ApJ...861..113D, 2011PhRvD..83b4024B} observations can help to break this degeneracy.
}

\vspace{0.2cm}

\noindent
A stochastic GW signal, either of astrophysical (foreground) or cosmological (background) origin, can be described in terms of the dimensionless energy density per logarithmic unit of frequency,
\begin{equation} \label{eq:omega_gw}
    \hat{\Omega}_{\rm GW}(f) = \frac{1}{\rho_c} \frac{d\rho(f)}{d\ln f} = \frac{2\pi^2}{3 H_0^2} f^2 h_c^2(f) .
\end{equation}
Here, $\rho_c$ is the critical density of the Universe, $\rho(f)$ is the energy density in GWs at frequency $f$, and 
$H_0=67.4\,\mathrm{km}\,\mathrm{s}^{-1}\,\mathrm{Mpc}^{-1}$ is the Hubble constant \cite{2020A&A...641A...6P}.
As shown in equation \ref{eq:omega_gw}, $\hat{\Omega}_{\rm GW}(f)$ is also related to the characteristic strain, $h_c(f)$, commonly used in the GW literature.

\begin{figure}[t]
    \begin{center}\includegraphics[width=0.7\textwidth]{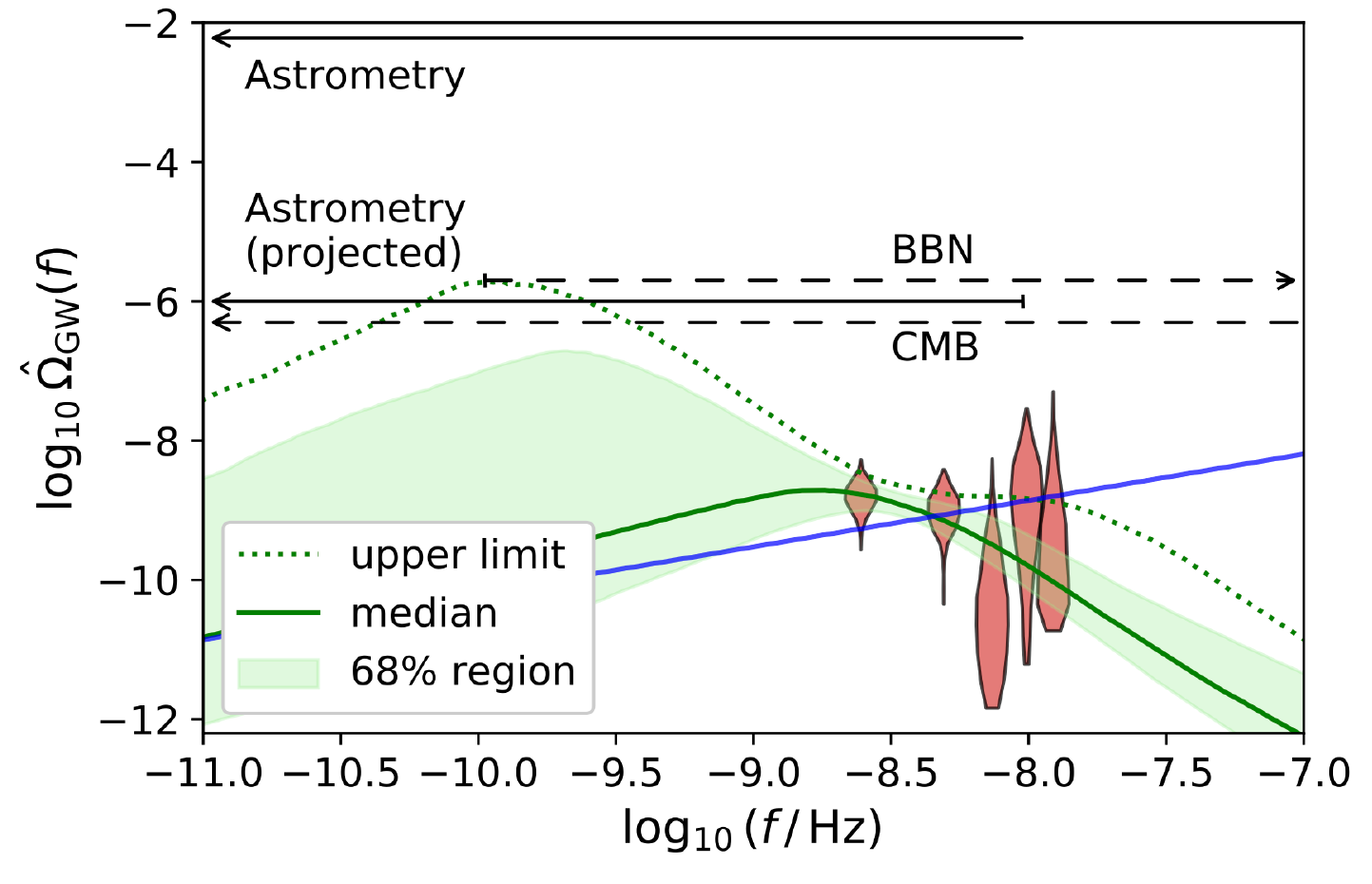}\end{center}
    \vspace{-0.7cm}
    \caption{
        \textbf{Combined constraints on GW backgrounds and foregrounds.}
        Red histograms show NANOGrav posteriors (in the 5 lowest frequency bins) on the energy density $\hat{\Omega}_{\rm GW}(f)$, see equation \ref{eq:omega_gw}.
        The blue line shows the median spectrum $\hat{\Omega}_{\rm GW}(f)$ obtained from the NANOGrav analysis assuming a BH astrophysical foreground origin for the signal with $\hat{\Omega}_{\rm GW}(f)\propto f^{2/3}$.
        The green line shows our median reconstructed spectrum for $\hat{\Omega}_{\rm GW}(f)$ based on the PTA-only data, assuming a cosmological PT background (for similar posteriors on the characteristic strain and the PTA \emph{delay}, see the Extended Data section).
        The shaded region (dotted line) indicates the $\pm 1\,\sigma$ uncertainty ($95\%$ upper limit) on the PT spectrum.
        Horizontal arrows show integrated bounds on the total energy $\Omega_{\rm GW}$ (see equation \ref{eq:Omega_gw}; solid and dashed lines indicate direct and indirect constraints respectively) with the horizontal extent of the arrows indicating the frequency range of integration.
    }
    \label{fig:Om_plot}
\end{figure}

PTAs directly measure $\hat{\Omega}_{\rm GW}(f)$ in the frequency range $1-100\,\mathrm{nHz}$ using a network of millisecond pulsars as extremely stable clocks, the arrival time of whose ``ticks'' are ever so slightly perturbed by GWs~\cite{1978SvA....22...36S, 1979ApJ...234.1100D, 1990ApJ...361..300F}.
Several collaborations have now undertaken a twenty year campaign of radio observations of $\sim 100$ of the most stable millisecond pulsars, resulting in a steadily improving sensitivity to GWs \cite{2015Sci...349.1522S, 2015MNRAS.453.2576L, 2016MNRAS.458.1267V, 2018ApJ...859...47A}.
The NANOGrav collaboration recently reported tentative evidence for a common stochastic process among the 45 millisecond pulsars in their $12.5\,\mathrm{year}$ data set consistent with an astrophysical GW background with $\hat{\Omega}_{\rm GW}(f_{\rm yr})=2.6-9.8\times 10^{-9}$ where $f_{\rm yr}=\mathrm{year}^{-1}$ [ref.~\cite{2020ApJ...905L..34A}]. PTA measurements are discussed further in Methods section \ref{sec:pulsar_timing}.
Although this is not yet a clear detection of GWs and the nature of the signal is still open to debate, the analyses presented in this letter will assume that it is indeed a GW signal. 
The observed signal is consistent both with an astrophysical foreground from massive ($\sim 10^6 - 10^9 \,M_\odot$) BH binaries~\cite{2021MNRAS.502L..99M} -- with $\hat{\Omega}_{\rm GW}(f)\propto f^{2/3}$ in the PTA band (blue line in Fig.~\ref{fig:Om_plot}), see Methods section~\ref{app:BBH_PT} -- and a red spectrum from a cosmological background generated by, for example, primordial BHs, cosmic strings, or dark-sector phase transitions see, for example refs.~\cite{
PhysRevLett.126.041303, 
PhysRevLett.126.051303, 
PhysRevLett.126.041304, 
PhysRevLett.126.041305, 
2021PhLB..81636238N,    
2021ScPP...10...47R,    
2021SCPMA..6490411A,    
PhysRevLett.115.181101}.

One can also observe the \emph{integrated} GW spectrum, i.e. the total energy density across a wide range of frequencies,
\begin{equation} \label{eq:Omega_gw}
    \Omega_{\rm GW} = \int_{f_{\rm min}}^{f_{\rm max}}\frac{\mathrm{d}f}{f}\; \hat{\Omega}_{\rm GW}(f).
\end{equation}  
Several indirect probes independently place constraints at the level $\Omega_{\rm GW} \lesssim 10^{-6}$ over different ultra-low frequency ranges (dashed black arrows in Fig.~\ref{fig:Om_plot}).
The energy in a GW background affects the growth of density perturbations as well as the cosmological expansion rate at the time of decoupling.
Therefore, CMB observations from WMAP or Planck, in combination with other large-scale structure surveys, are able to place integrated constraints on $\Omega_{\rm GW}$ [ref.~\cite{2020JCAP...10..002C, 2006PhRvL..97b1301S}] with $f_{\rm min}\sim 10-100\, f_{\rm Hubble}$, where $f_{\rm Hubble}=H_0$.
Elsewhere, the success of big bang nucleosynthesis (BBN) predictions for the primordial abundances of light elements constraints extra, unbudgeted energy (including GWs) at the time of nucleosynthesis. 
As BBN occurs before recombination, at a temperature $T_{\rm BBN}\sim 1\,\mathrm{MeV}$, this only constrains the GW spectrum above $f_{\rm min} = f_{\rm BBN}\sim 10^{-10}\,\mathrm{Hz}$ [ref.~\cite{2014ApJ...781...31C, Maggiore:2018sht}].
Both the CMB and BBN bounds probe GWs in the early Universe and are insensitive to those produced at later cosmic times by astrophysical processes.
Furthermore, both bounds depend on assumptions such as the density distribution of perturbations and the effective number of neutrino species, and probe the GW background only indirectly; these early Universe constraints are reviewed in Methods section \ref{subsec:cosmo}.

Astrometry is the only other technique directly sensitive to ultra-low frequency GWs at a level potentially competitive with the aforementioned probes.
Photons from extragalactic sources, following null geodesics in the background metric, are deflected by ultra-low GWs causing changes in the apparent positions (i.e. proper motions) of distant objects.
Very long baseline interferometry (VLBI) observations of the proper motions of a few hundred quasars taken over several years place an upper limit of $\Omega_{\rm GW}\lesssim 6\times 10^{-3}$ [ref.~\cite{2018ApJ...861..113D}] on GWs below a frequency set by the span of observations, $f_{\rm max}\sim (3 \,\mathrm{year})^{-1}$.
The projections are that this might be improved to as low as $\lesssim 10^{-6}$ by the end of the Gaia mission~\cite{2011PhRvD..83b4024B} using a greatly expanded catalog of quasars.
The astrometric bounds are plotted using solid arrows in Fig.~\ref{fig:Om_plot}.
In contrast to CMB and BBN bounds, astrometric observations directly constrain GWs in the local Universe, which includes any produced by astrophysical processes after recombination, and distinguishes them from other forms of energy (astrometry even distinguishes between different possible GW polarisations~\cite{PhysRevD.97.124058}).
Astrometric observations are also reviewed in Methods section \ref{subsec:astro}.

We assess the importance of including data that provide information on the integrated spectrum, $\Omega_\mathrm{GW}$, together with PTA data that measure the frequency dependent spectrum, $\hat\Omega_\mathrm{GW}(f)$.
We do this by considering several example model spectra, many of which contain significant power below the low frequency reach of PTAs, but are nevertheless subject to the integrated constraints (see Fig.~\ref{app:BBH_PT}). We consider these models to be proxies for a wide range of possible exotic physics. 
Jointly analysing these data sets maximises the ability to distinguish between the overlapping contributions from astrophysical foregrounds and cosmological backgrounds and is vital to interpreting any signal that may be observed. 

\begin{figure}[t]
    \begin{center}\includegraphics[width=0.8\textwidth]{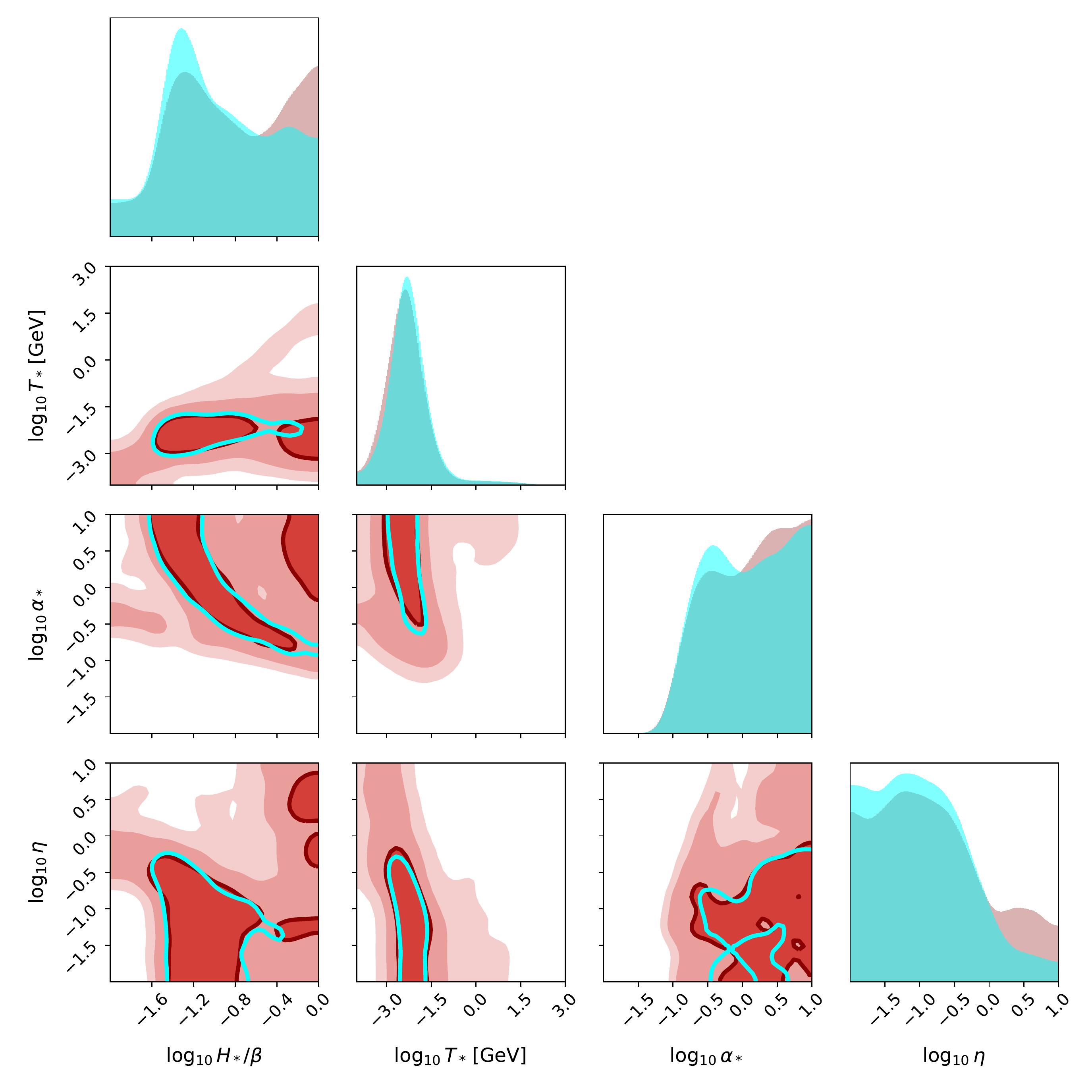}\end{center}
    \vspace{-0.7cm}
    \caption{
        \textbf{Posteriors on the phase transition model parameters.} 
        The 4 parameters are the temperature, $T_*$, and strength, $\alpha_*$, of the PT, the bubble nucleation rate, $\beta/H_*$, in units of the Hubble parameter at the transition, and the dimensionless friction parameter, $\eta$.
        Posteriors obtained from the PTA data alone are shown in red with the 50\% contour indicated. 
        Shown in light blue is the 50\% contour for the posterior obtained with the additional integrated constraint $\Omega_{\rm GW}<10^{-6}$.
    }
    \label{fig:corner_plot}
\end{figure}

We start by considering a particular phase transition (PT) model~\cite{2009JCAP...12..024C, 2017PhRvD..95b4009J, 2017PhRvD..96j3520H} which is here determined to occur in the temperature range $\approx 1-300\,\mathrm{MeV}$.
In order to be consistent with standard model physics at these low collider energies, the transition must occur in a \emph{dark sector}, decoupled from the standard model.
GWs are generated during the PT by collisions of bubbles and sound waves, as well as turbulence.
The PT spectrum fitted only to the NANOGrav data (green line in Fig.~\ref{fig:Om_plot}) allows for peaks below $\sim 1\,\mathrm{nHz}$ and (at $85\%$ confidence) for $\Omega_{\rm GW}>10^{-6}$.
As has been pointed out\cite{NANOGrav_PT}, PTA measurements alone are currently unable to distinguish this background from an astrophysical foreground.
However, including integrated bounds from CMB, BBN and astrometry further constrains the PT model. 
In Methods section \ref{sec:results} we describe our analysis of the NANOGrav data using the PT of ref \cite{NANOGrav_PT}, both with and without the inclusion of an integrated bound, $\Omega_{\rm GW}<10^{-6}$. 
Although the integrated bound is insufficiently constraining to completely rule out the PT model, we find it does eliminate a significant fraction of the PT model parameter space posterior probability volume; see Fig.~\ref{fig:corner_plot}.
To quantify this improvement, we calculate the squared Hellinger distance ($0\leq d_{\rm H}^2 \leq 1$) between the two distributions in Fig.~\ref{fig:corner_plot} to be $d_{\rm H}^2=0.07$. 
This modest improvement can be interpreted as integrated constraint ruling out 14\% of the PT parameter space posterior volume compared to the PTA-only analysis; in particular, regions of large $\eta$, $\alpha_*$ and $H_* / \beta$ which predict $\hat{\Omega}_{\rm GW}(f)$ spectra that peak strongly below $\sim 1\mathrm{nHz}$ are ruled out.
This can be seen most clearly by comparing the two $50\%$ contours in Fig.~\ref{fig:corner_plot}. 
The combined results slightly improve the constraint on the temperature of the transition to be in the range $T_*\sim 1-300\,\mathrm{MeV}$ (10\% and 90\% quantiles).

\begin{figure}[t]
    \centering
    \begin{minipage}{0.49\textwidth}
        \centering
        \vspace{1.8cm}
        \includegraphics[width=\textwidth]{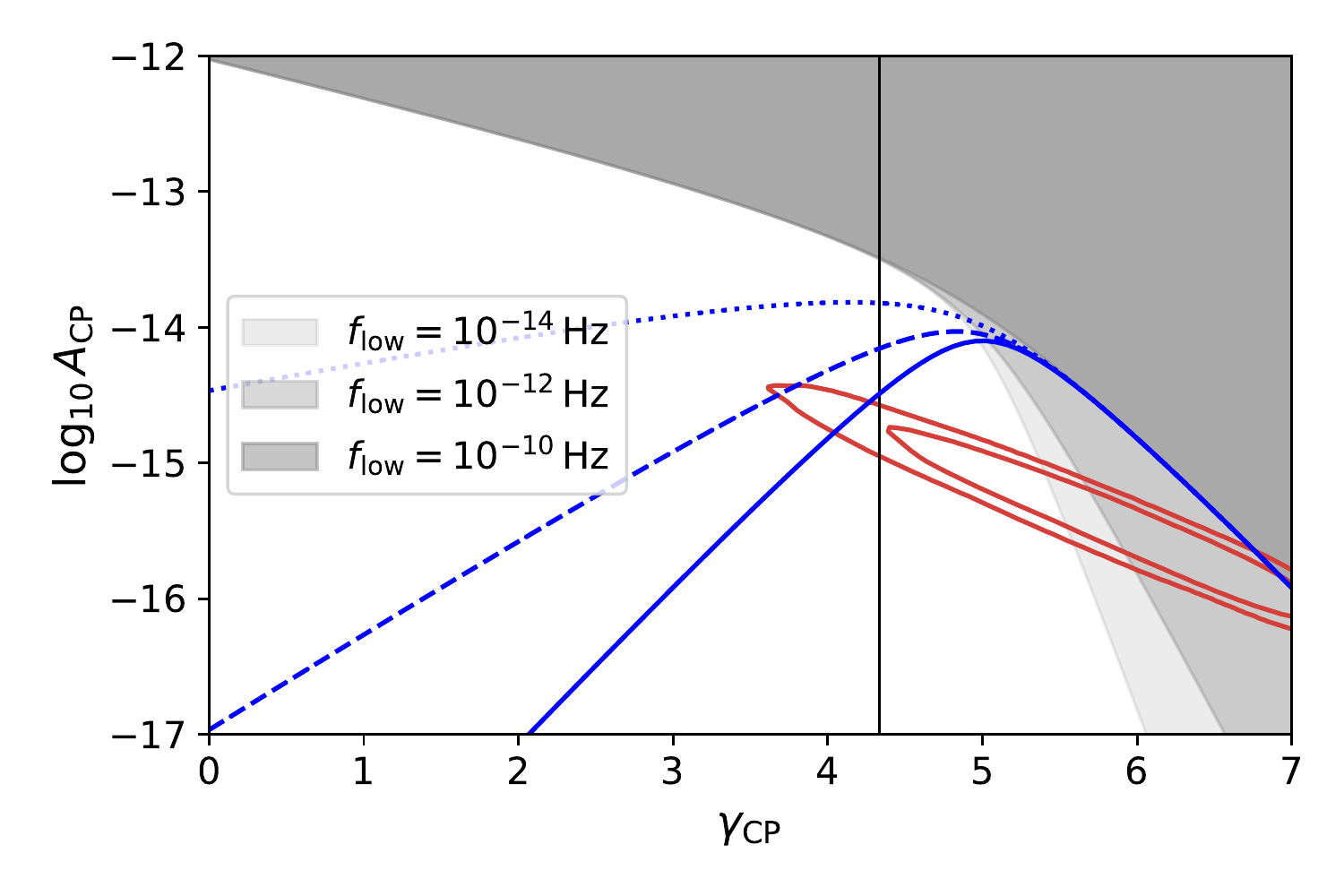}
        \vspace{0.5cm}
        \caption{\textbf{Combined constraints on power law stochastic GW signals.}
        Red contours (50\% and 90\%) indicate PTA-only posteriors on the power-law spectrum
        amplitude and slope parameters.
        The vertical line indicates the astrophysical prediction, $\gamma_{\rm CP}=13/3$.
        The incorporation of an integrated constraint, $\Omega_{\rm GW}<10^{-6}$, helps to rule out some redder spectra (i.e. those with larger $\gamma_{\rm CP}$), although this depends on the low frequency cutoff, $f_{\rm low}$, of the spectra; the shaded regions show the excluded regions for 3 choices of $f_{\rm low}$.
        The integrated constraint also helps to rule out extremely blue spectra (although these are anyway mostly incompatible with the PTA measurements) as shown by the constraints plotted in blue which depend on a high frequency cutoff in the spectra; results are shown for $f_{\rm high}=10^{-5}\,\mathrm{Hz}$ (solid), $f_{\rm high}=10^{-6}\,\mathrm{Hz}$ (dashed), and $f_{\rm high}=10^{-7}\,\mathrm{Hz}$ (dotted).\label{fig:PL_plot}}
    \end{minipage}\hfill
    \begin{minipage}{0.49\textwidth}
        \centering
        \includegraphics[width=\textwidth]{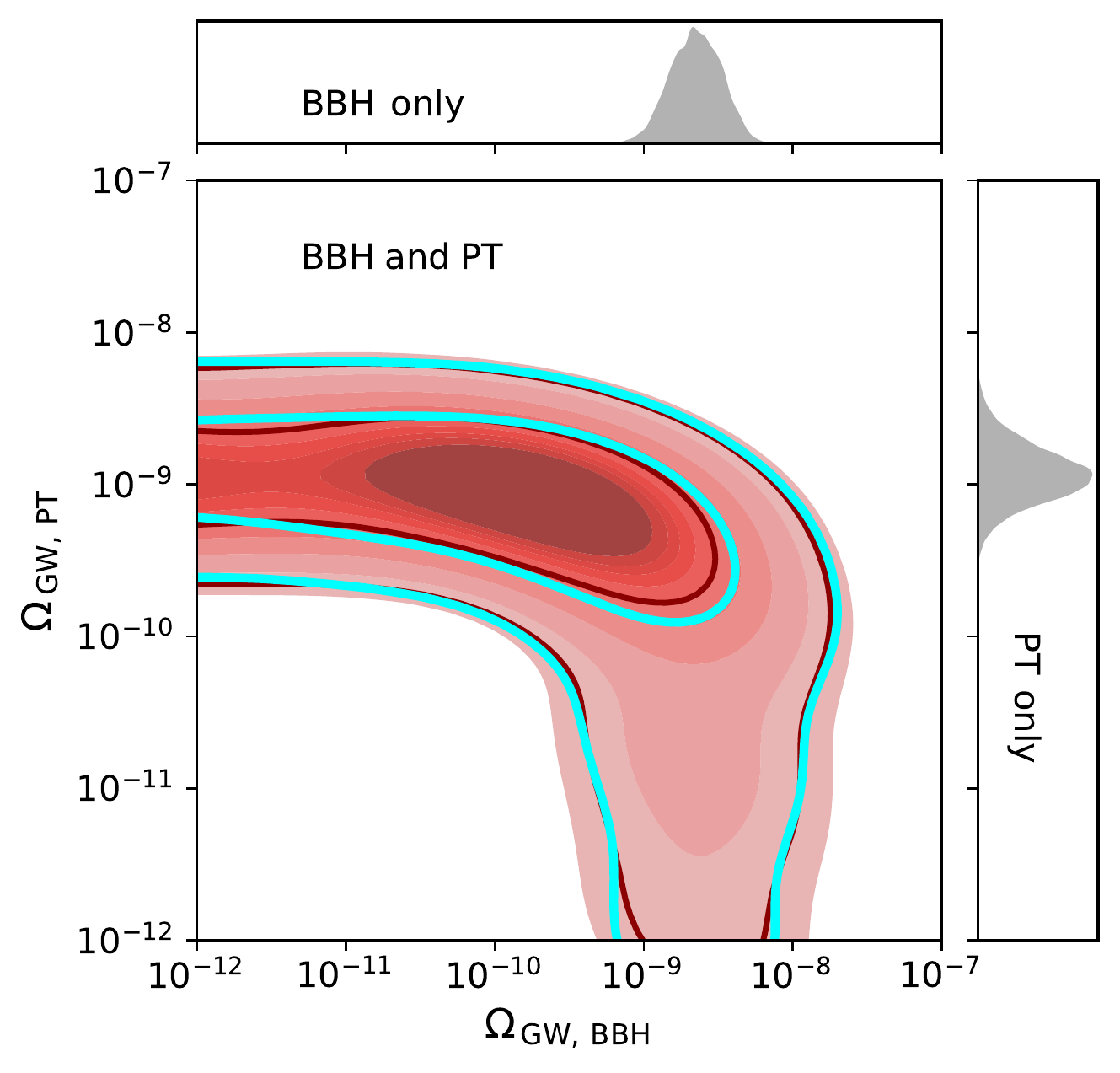}
        \vspace{-0.3cm}
        \caption{\textbf{Combined measurements of the GW energy densities in the background and foreground.}
        The top panel shows the PTA-only posterior on the energy density $\Omega_{\mathrm{GW},\,\mathrm{BBH}}$ assuming the signal comes from only BBHs.
        The right panel shows the posterior on $\Omega_{\mathrm{GW},\,\mathrm{PT}}$ assuming the signal comes only from a PT. In both cases the energy density is relatively well constrained.
        The red histogram in the main plot shows the joint posterior on the two energy densities using a model that allows for arbitrary overlapping contributions from both BBHs and a PT.
        Note that the two side panels are \emph{not} 1-dimensional marginalised posteriors of the central distribution, they instead show the results of separate, lower-dimensional analyses.
        The cyan contours (50\% and 90\%) show the results when integrated low frequency constraints are incorporated into the analysis. \label{fig:BBH_PT}}
    \end{minipage}
\end{figure}

Additionally, we perform a separate analysis for a simple power-law GW spectrum model,
\begin{equation}
    h_c(f) = A_{\rm CP} \left(\frac{f}{f_{\rm yr}}\right)^{\alpha_\textsc{CP}}, \quad \mathrm{for} \quad f_{\rm low}<f<f_{\rm high}.
\end{equation}
where $\alpha_{\rm CP} = (3-\gamma_{\rm CP})/2$.
The NANOGrav results alone are consistent with a range of spectral slopes that allow for a variety of possible early-Universe interpretations.
In fact, the NANOGrav data favours a spectrum redder (i.e. larger $\gamma_{\rm CP}$) than the astrophysical prediction of $\gamma_{\rm CP}=13/3$ (see Fig.~\ref{fig:PL_plot}).
We find that the inclusion of an integrated constraint remains consistent with an astrophysical interpretation of the signal but helps to rule out extremely red spectra, although the exact constraint is sensitive to the placement of the low frequency cutoff of the spectrum. 

We also report results of analyses that model the astrophysical foreground from binary BHs, possibly overlapping with a PT cosmological background.
The details of the astrophysical model are given in Methods section \ref{app:BBH_PT}.
If the stochastic GW signal is assumed to be entirely from binary BHs, then the total GW energy density in the astrophysical foreground in the PTA frequency band can be measured from the NANOGrav data; the posterior is shown in the top panel of Fig.~\ref{fig:BBH_PT}.
Alternatively, if the signal is assumed to be entirely from a PT, then the same GW energy density in the cosmological background can be measured; the posterior is shown in the right panel of Fig.~\ref{fig:BBH_PT}.
Finally, if the analysis allows the signal to be modelled as an arbitrary combination of an astrophysical foreground from binary BHs superposed on a cosmological PT background then the 2-dimensional red posterior in the main panel of Fig.~\ref{fig:BBH_PT} reveals the difficulty in separating the two signal components.
If the analysis is repeated with the inclusion of the integrated constraint (cyan contour in Fig.~\ref{fig:BBH_PT}) then an further 10\% of the posterior probability volume is ruled out, but the degeneracy remains.
The details of these analyses are given in Methods section \ref{app:BBH_PT}.

In summary, with a detection imminent \cite{2021ApJ...911L..34P}, it is necessary to separate clearly the astrophysical foreground from a variety of possible cosmological backgrounds in order to reliably interpret an ultra-low frequency GW signal.
This is difficult with PTAs alone due to their limited ability to measure the spectral slope and/or shape [longer term, the IPTA \cite{2019MNRAS.490.4666P} and SKA, e.g.  \cite{2015aska.confE..37J} will increase this ability].
The situation can be improved by consistently combining information from other probes of the ultra-low frequency GW background that place constraints on the integrated spectrum.
These include indirect cosmological probes sensitive to GWs (and other relativistic species) at high redshifts as well as astrometric observations that directly constrain GWs produced across all cosmic times.
It is therefore important to improve as far as possible these bounds, in particular the direct astrometric bound that can be placed with Gaia in combination with VLBI.

\section*{Acknowledgements} AV acknowledges the support of the Royal Society and Wolfson Foundation.
We thank Wolfram Ratzinger and Kai Schmitz for useful discussions on the analysis.

\section*{Contributions} Both C.J.M.\ and A.V.\ jointly conceived this study. C.J.M.\ performed the numerical calculations. Both C.J.M.\ and A.V.\ jointly analysed and interpreted the results and wrote the manuscript.

\section*{Competing Interests} The authors declare that they have no competing financial interests.

\section*{Data Availability} The data used for the analyses presented is publically available from \url{https://data.nanograv.org}.

\section*{Code Availability} The code used for the analyses presented here, including that to make all figures is avaiable at \url{https://github.com/cjm96/PTA_and_IntegratedConstraints}
which uses code from \url{https://github.com/AMitridate/bokeh/tree/master/bokeh-app}.

\section*{Correspondence} Correspondence and requests for materials should be addressed to C.J.M. cmoore@star.sr.bham.ac.uk


\section*{Methods}

\section{PTA Measurements of \ensuremath{\hat{\Omega}_{\rm GW}(\lowercase{f})}}\label{sec:pulsar_timing}

Measurements of the arrival time of pulses from an ensemble of millisecond pulsars provide a means to directly detect gravitational waves (GWs) in the nanohertz frequency window~\cite{1978SvA....22...36S, 1979ApJ...234.1100D, 1990ApJ...361..300F}.
A PTA uses several large radio telescopes to make such measurements every few weeks 
over a decades long period.
A GW stochastic signal affects the time of arrival of the pulses from each of the pulsars in the array with a characteristic spatial correlation that depends on each pulsar pair's angular separation in the sky described by the so-called Hellings and Downs correlation \cite{1983ApJ...265L..39H}. 

A planar GW from a source in direction $\hat{q}$ causes a redshift in a pulsars pulse-frequency according to
\begin{equation}
    z(t, \hat{q}) = \frac{1}{2} \frac{\hat{p}^i\,\hat{p}^j}{1 - \hat{q} \cdot \hat{p}} \Delta h_{ij}(t),
\end{equation}
where $\hat{p}$ is the direction to the pulsar, and $\Delta h_{ij}(t)$ is the difference in spatial metric perturbations at the Earth (or solar system barycentre) at time $t$ and the pulsar at retarded time $t-D/c$, where $D$ is the distance to the pulsar and $c$ is the speed of light.
The actual measurable quantity in pulsar timing is the delay in pulse arrival time, or timing residual, given by
\begin{equation}
    r(t) = \int^t dt^\prime\; z(t^\prime).
\end{equation}
For an isotropic, stochastic signal, the timing residuals have a power-cross spectrum between pulsar $a$ and $b$ given by
\begin{equation}\label{eq:S_f}
    S_{ab}(f) = \Gamma_{ab} \frac{A_{\rm CP}^2}{12\pi^2} \left(\frac{f}{f_{\rm yr}}\right)^{-\gamma} f_{\rm yr}^{-3}.
\end{equation}
Setting $\Gamma_{ab}=1$, as there is currently no evidence for inter-pulsar correlations, gives $S_{ab}(f)=S(f)$ as a function of frequency only, which can be related to $h_c(f)$ or $\hat{\Omega}_{\rm GW}(f)$ used in the main text.
Posteriors on the PT model spectra of all these related quantities are shown in Figs.~\ref{fig:Om_plot} and in the figure in the Extended Data section.

Several constraints exist on the amplitude of a stochastic background from the 3 large independent PTA collaborations: Parkes (PPTA~\cite{2015Sci...349.1522S}), the European PTA (EPTA~\cite{2015MNRAS.453.2576L}), and NANOGrav~\cite{2018ApJ...859...47A}.
These PTAs, and others, collaborate under the umbrella of the international PTA (IPTA) which has also placed an upper limit on the GW background~\cite{2016MNRAS.458.1267V}.

Recently, NANOGrav~\cite{2020ApJ...905L..34A} reported tentative evidence for a stochastic common process (CP) among the 45 millisecond pulsars in their 12.5 year data set.
Assuming this is due to an astrophysical GW foreground from supermassive BH binaries with the expected $\gamma_{\rm CP}=13/3$ slope \cite{2008MNRAS.390..192S, 2013MNRAS.433L...1S, 2017MNRAS.471.4508K}, the NANOGrav data are consistent with a GW signal with amplitude $h_c (f_{\rm yr}) = 1.37-2.67\times 10^{-15}$ (5\%--95\%; using equation \ref{eq:omega_gw}, this translates to the range on $\hat{\Omega}_{\rm GW}(f_{\rm yr})$ stated in the main text) \cite{2020ApJ...905L..34A}.
However, the observations currently do not show conclusive evidence for the inter-pulsar Hellings and Downs correlations; without this, the result cannot yet be regarded as a clear detection of GWs.

In our analysis, instead of the fiducial astrophysical model, we focus on the results of the NANOGrav free-spectrum analysis where the signal in each frequency bin is fitted separately. 
The most constraining measurements come at the lowest frequencies.
As was done in \cite{2020ApJ...905L..34A}, we use the posteriors on the first 5 frequency bins. 
These measurements are plotted in Fig.~\ref{fig:Om_plot} (and in the Extended Data section) where it can be seen that the data are consistent with a wide range of GW spectral shapes which allows for a variety of non-standard interpretations, including the first-order dark-sector PTs considered here \cite{2021PhLB..81636238N, 2021ScPP...10...47R,2021SCPMA..6490411A, PhysRevLett.115.181101}, but also QCD PTs \cite{2021arXiv210212428B, PhysRevD.103.L041302}, topological defects \cite{PhysRevLett.126.041304, PhysRevLett.126.041305, 2021arXiv210108012L, 2021JCAP...06..048A}, primordial BHs \cite{PhysRevLett.126.041303, PhysRevLett.126.051303} and inflation \cite{Vagnozzi:2020gtf}.

\section{Integrated Constraints on \ensuremath{\Omega_{\rm GW}}}\label{sec:integrated}

\subsection{Cosmological~--~}\label{subsec:cosmo}
Cosmologically, a GW background behaves as a free-streaming gas of massless particles, in exactly in the same way as do massless neutrinos. It therefore will have a similar effect on the growth of density perturbations in addition to its effect on the CMB via the expansion rate, including at the time of decoupling. 

One then needs to consider the density distribution of the perturbations. This is related to the choice of an \textit{adiabatic} or \textit{homogeneous} spectrum of density perturbations.
If the energy-density perturbations have the same density distribution as the other relativistic species they are said to be \textit{adiabatic}. 
In this case the effects on the CMB, large-scale structure and baryon acoustic oscillations are indistinguishable from those of massless neutrinos. 
However, it is the cumulative effect of all relativistic species that is measured from CMB and large-scale structure observations.
This is commonly parameterised in the literature as an effective number of relativistic species, $N_\mathrm{eff}$. 

Assuming adiabatic density perturbations, and using the third year WMAP data release (in combination with other small-scale CMB measurements, constraints from the Lyman-$\alpha$ forest and the Sloan Digital Sky Survey),
Ref~\cite{2006PhRvL..97b1301S} obtains $\Omega_{\rm GW}\,h_{100}^2 \le 4.0\times 10^{-5}$ at 95\% confidence ($h_{100}$ is the present day Hubble constant in units of $100\,\mathrm{km} \,\mathrm{s}^{-1}\,\mathrm{Mpc}^{-1}$). 
Using data from the Planck mission, ref~\cite{2020JCAP...10..002C} updated this bound and obtained $\Omega_{\rm GW}h_{100}^2\le 1.7\times 10^{-6}$ (at 95\% confidence).
This integrated bound applies over a frequency range above $f_{\rm min}\sim 100\,f_{\rm Hubble} \sim 10^{-15}\,\mathrm{Hz}$.

However, if the density perturbations are not adiabatic, this limit does not apply. 
If one instead assumes that the perturbations are initially homogeneous, then GW perturbations evolve differently to the neutrino perturbations and the degeneracy is broken. 
This modifies the growth of perturbations, especially at large scale.
If the background is produced by some physical mechanisms that leaves its primordial density uncorrelated with the curvature perturbations -- e.g. phase transitions, and anything post-inflation -- then this alternate limit applies. 
Using the WMAP data Ref~\cite{2006PhRvL..97b1301S} obtain $\Omega_{\rm GW}\,h_{100}^2 \le 8.4\times 10^{-6}$ at 95\% confidence, which is \textit{tighter} than under the adibatic assumption.
This bound has been revised using recent Planck data to yield $\Omega_{\rm GW}\,h_{100}^2 \le 2.9\times 10^{-7}$ by Ref~\cite{2020JCAP...10..002C}.
This alternate bound also applies over a frequency range above $f_{\rm min}\sim 100\,f_{\rm Hubble} \sim 10^{-15}\,\mathrm{Hz}$.

Big bang nucleosynthesis (BBN) bounds on $\Omega_{\rm GW}$ come from the measured primordial abundance of light atomic nuclei.
The theoretical prediction of the abundances of these elements (including D, ${}^{3}$He, ${}^{4}$He, and ${}^{7}$Li) is among the most impressive achievements of modern concordance cosmology.
The bounds on $\Omega_{\rm GW}$ come from the dependence of the abundances on the expansion rate of the universe at the time of BBN which occurs at at a temperature $T_{\rm BBN}\sim 1\,\mathrm{MeV}$.
This bound constrains GWs present at the time of nucleosynthesis (i.e. only those GW generated at earlier cosmic times).
Refs~\cite{2014ApJ...781...31C, Maggiore:2018sht} find a bound of $\Omega_{\rm GW}\,h_{100}^2\le 1.3\times 10^{-6}$. Although this bound also is also coupled with the effective number of neutrinos.
The minimum frequency over which the integrated BBN bound applies is set by the horizon scale at the time of BBN, redshifted to the present day. 
Taking $T_{\rm BBN}\sim 1\,\mathrm{MeV}$, this gives $f_{\rm min}=f_{\rm BBN}\sim 10^{-10}\,\mathrm{Hz}$, although, in reality, there is a smooth transition over a range of frequencies $\sim 10^{-12} - 10^{-10}\,\mathrm{Hz}$.
For the PT models considered here, the frequency range over which the BBN bound is obtained is less constraining than that for the CMB bound. This is because the PT models can have can have a large fraction of their total energy at frequencies $f<f_{\rm BBN}$, thereby partially evading the BBN constraint (see arrows in Fig.~\ref{fig:Om_plot}).

For both the the CMB and BBN bounds, the upper limit of frequency sensitivity is effectively $f_{\rm max}\rightarrow\infty$ as the only known limit on the GW wavelength is set by Planck-scale physics at the relevant cosmological time.

\subsection{Astrometric~--~}\label{subsec:astro}
The idea that GWs cause observable astrometric effects in optical sources is an old one \cite{1990NCimB.105.1141B, 1997ApJ...484..545K} but the prospects for a detection are generally not considered promising \cite{2010IAUS..261..234S}.
For a distant optical source, such as a quasar (QSO), in direction $n^i$, the GW induced astrometric deflection $\delta n^i$ depends on the local GW metric perturbation in the solar system, $h_{ij}$, caused by a GW source in direction $q^i$. The deflection is given by Ref~\cite{1996ApJ...465..566P} (see also~\cite{2011PhRvD..83b4024B, PhysRevD.97.124058})
\begin{equation} \label{eq:GW_astro_response}
  \delta n^i(t) = \bigg[\frac{(n^i-q^i)n^jn^k}{2(1-q_l n^l)} - \frac{1}{2}\delta^{ij} n^k\bigg]h_{jk}(t) .
\end{equation}

Early efforts to detect this effect used very long baseline interferometry (VLBI) \cite{SHAPIRO1976261} measurements of the positions of radio sources.
However, interest increased following the launch of 
Gaia~\cite{2016A&A...595A...1G} 
and the early data releases. 

As an example, consider the Gaia eDR3 release \cite{2021A&A...649A...1G} which has a duration of {$34\,$months$\,\approx \!10^{8}\,$s}. 
There are two distinct frequency regimes in which GWs can potentially be detected.
In the range ${f\gtrsim 10^{-8}\,\mathrm{Hz}}$, Gaia can look for periodic effects of GWs with periods shorter than the mission duration; see, for example~\cite{2017PhRvL.119z1102M, 0264-9381-35-4-045005}. 
Alternatively, Gaia can search for the secular effects of ultra-low frequency GWs with periods longer than the mission duration; see, for example~\cite{2012A&A...547A..59M}.
It is the later case that is our focus here.
These ultra-low frequency GWs produce constant (rather than periodic) apparent velocities of distant objects across the sky, known as proper motions.
Unlike PTA observations which directly constrain the GW spectrum $\hat{\Omega}_{\rm GW}(f)$, astrometric observations taken over time span $T$ place constraints on the integrated spectrum $\Omega_{\rm GW}$ with $f_{\rm max}\sim 1/T$.
The lower frequency sensitivity is only limited by the Hubble scale, $f_{\rm min}\sim f_{\rm Hubble}$.

As in pulsar timing, the \emph{smoking gun} signature of stochastic GW background would be a distinctive correlation in the pattern of QSO proper motions. 
This correlation pattern is predominantly quadrupolar (i.e.\ $\ell=2$) but with subdominant contributions from higher harmonics \cite{2011PhRvD..83b4024B} [see also Refs~\cite{PhysRevD.97.124058, 2019PhRvD..99f3002Q} for a discussion using an alternative formalism] and can be thought of as the astrometric analog of the PTA Hellings and Downs curve.
This correlation pattern is an extremely robust prediction within the theory of general relativity as it depends on no free parameters.
Measuring this correlation pattern can be a powerful test of the theory as it is sensitive to both the polarisation content of the GW background and the propagation speed of the GWs \cite{PhysRevD.97.124058,2018PhRvD..98b4020O, 2020PhRvD.101b4038M}.

Early astrometric constraints on the GW background came from VLBI.
Ref~\cite{1997ApJ...485...87G} used measurements of 323 extragalactic sources to place the constraint $\Omega_{\rm GW}\lesssim 10^{-1}$ with $f^{-1}_{\rm max}\sim  10\,\mathrm{year}$.
This was subsequently improved~\cite{2011A&A...529A..91T} to $\Omega_{\rm GW}\lesssim 4\times 10^{-3}$ with $f_{\rm max}^{-1}\sim 20\,\mathrm{year}$ using 555 sources.
Both analyses used only the quadrupolar component of the proper motions.

Prior to launch, several studies attempted to forecast the improvement on this bound that might be expected by the end of the Gaia mission.
Gaia will observe a much larger catalog of several million QSOs, albeit with larger proper motion errors than obtained with VLBI.
Ref~\cite{2011PhRvD..83b4024B} projected a constraint of $\Omega_{\rm GW}\lesssim 10^{-6}$ with $f_{\rm max}^{-1}\sim 1\,\mathrm{year}$.
A more conservative assessment~\cite{2012A&A...547A..59M} projected $\Omega_{\rm GW}\lesssim 8\times 10^{-5}$ with $f^{-1}_{\rm max}\sim 5\,\mathrm{year}$.

More recently, Ref~\cite{2018ApJ...861..113D} placed an updated VLBI constraint of $\Omega_{\rm GW}\lesssim 6\times 10^{-3}$ with $f^{-1}_{\rm max}\sim 20\,\mathrm{year}$ using 713 radio sources and $\Omega_{\rm GW}\lesssim 10^{-2}$ with $f_{\rm max}^{-1} \sim 1\,\mathrm{year}$ using Gaia DR1 in combination with a smaller set of 508 VLBI sources.
Ref~\cite{2018ApJ...861..113D} also used primarily the quadrupole component of the proper motions in their analysis, but also considered octupole contributions.

Further improvements towards the limit of $\Omega_{\rm GW}\lesssim 10^{-6}$ projected by \cite{2011PhRvD..83b4024B} might be expected due to: (i) longer baselines out to $\approx 10\,\mathrm{years}$ in future Gaia data releases, (ii) larger data sets with more QSOs and better source selection, (iii) reduced astrometric errors and improved control over systematics, (iv) consistent inclusion of higher harmonics in the analysis, (v) simultaneous inclusion of different astrometric datasets obtained over different baselines and that therefore constrain the integrated spectrum over different frequency bandwidths (extra care will be required if datasets are not independent, as would be the case with different Gaia data releases). However, see also \cite{2020ApJ...890..146P} who point out some of difficulties that need to overcome to maximise the astrometric GW sensitivity.
Longer term, future astrometric observatories \cite{2016arXiv160907325H, 2017arXiv170701348T} might also place tighter constraints.

\section{Main Analyses}\label{sec:results}

We use the NANOGrav $T=12.5\,\mathrm{year}$ posteriors~\cite{2020ApJ...905L..34A} on the delay, $\sqrt{S(f_i)/T}$, at the 5 lowest frequencies, $f_i$.
These posterior are from the free spectrum analysis in which the amplitude of the GW spectrum in each frequency bin is allowed to vary independently.
We use the public data products from \url{https://data.nanograv.org}.
These posteriors were converted to $\hat{\Omega}_{\rm GW}(f_i)$ via equations \ref{eq:omega_gw} and \ref{eq:S_f} and used to build kernel density estimates (KDEs) for the posterior at each frequency.

We consider three models for the spectrum, $\hat{\Omega}_{\rm GW}(f)$. First, the physically motivated PT model from \cite{2009JCAP...12..024C, 2017PhRvD..95b4009J, 2017PhRvD..96j3520H}, for which we use the semi-analytic parameterisation described in \cite{NANOGrav_PT}. 
This contains 4 free parameters: the temperature, $T_*$, and strength, $\alpha_*$, of the transition, the dimensionless nucleation rate $\beta/H_*$, and the dimensionless friction parameter, $\eta$. For each of these parameters we use log-uniform priors over the ranges shown in Fig.~\ref{fig:corner_plot}.
Second, the simple power-law model which is described by just an amplitude, $A_{\rm CP}$, and spectral slope, $\gamma_{\rm CP}$. Uniform (log-uniform) priors on $\gamma_{\rm CP}$ ($A_{\rm CP}$) over the ranges shown in Fig~\ref{fig:PL_plot} were used. 
The power-law model also requires low and high frequency cutoffs; these were fixed during inference runs and results are shown for several choices.
Finally, we also consider an astrophysical model for the spectrum from massive binary BHs; this is described in Methods section \ref{app:BBH_PT}.

Our PTA-only log-likelihood function is obtained by evaluating the spectrum at the 5 values $f_i$ and summing the log-PDFs of the 5 independent KDEs at these values. 
If required, an integrated constraint is imposed by setting the log-likelihood to $-\infty$ if the integrated energy across all frequencies exceeds $\Omega_{\rm GW}>10^{-6}$.
Sampling was performed with the \textsc{dynesty} \cite{2020MNRAS.493.3132S} implementation of the nested sampling algorithm \cite{2004AIPC..735..395S}. 
Posteriors on the PT parameters are shown in Fig.~\ref{fig:corner_plot}, for the power-law parameters in Fig.~\ref{fig:PL_plot}.

It should be pointed out that our PTA-only analysis differs from that in \cite{NANOGrav_PT} where the $12.5\,\mathrm{year}$ data set was reanalysed in its entirety, which includes marginalising again over the red-noise parameters in each pulsar.
Here we exploit the posterior density functions on the 5 lowest frequency bins that have already been obtained from a previous full free spectrum analysis that has already marginalised over the red-noise parameters. 
We use these posterior samples to perform a second Bayesian analysis on the GW background parameters only. 
In this sense, ours is a hierarchical analysis.
The agreement between the two approaches can be clearly seen by comparing Fig.~\ref{fig:corner_plot} with Fig.~4 in \cite{NANOGrav_PT}.
This hierarchical approach allows for an easy modification of the likelihood to include additional information, such as the integrated constraints considered here.
Our approach can therefore be used in future to analyse multiple datasets and explore a wider range of possible GW production mechanisms.

\section{Additional analyses of astrophysical foregrounds and cosmological backgrounds}\label{app:BBH_PT}

It is expected that a cosmological GW background will appear behind an astrophysical foreground generated by inspiralling binary black holes (BBHs) at late cosmic times.
This section presents the results of an analysis that allows for an arbitrary overlapping combination of the two sources.

The astrophysical prediction for the BBH foreground is for a power law spectrum with a specific value for the slope, $\gamma_{\rm CP}=13/3$ [refs.\cite{2008MNRAS.390..192S, 2013MNRAS.433L...1S, 2017MNRAS.471.4508K}]. 
This assumes that binaries are on circular orbits which evolve solely due to GW emission. 
We adopt the model for the BBH foreground presented in Ref~\cite{2016MNRAS.455L..72M} (and used in Ref~\cite{2021MNRAS.502L..99M}) which is briefly summarised here.

A population of BBHs is described by a distribution ${\rm d}^2 n / ({\rm d}z\, {\rm d}\!\log_{10} {\cal M}$): the number density of binaries per unit redshift, $z$, per unit logarithmic chirp mass, ${\cal M}$. 
(For a binary with individual masses $m_1$ and $m_2$, the chirp mass is defined as ${\cal M} = (m_1 m_2)^{3/5}/ (m_1+m_2)^{1/5}$.)
The characteristic amplitude, $h_c(f)$, which is related to the spectrum $\hat{\Omega}_{\rm GW}(f)$ via equation~\ref{eq:omega_gw}, is given by
\begin{equation}
    h_c^2(f) = \frac{4 G^{5/3}}{3 \pi^{1/3} c^2} f^{-4/3} \int {\rm d} \log_{10}{\cal M}\, \int {\rm d} z \;(1+z)^{-1/3} {\cal M}^{5/3} \frac{{\rm d}^2 n}{{\rm d}z\, {\rm d}\! \log_{10} {\cal M}}\,,
    \label{eqn:hc}
\end{equation}
where $G$ is Newton's constant.
Following Ref~\cite{2016MNRAS.455L..72M}, the integration is performed over the chirp mass range $10^6-10^{11}\,M_\odot$ and the redshift range $0-5$.
Also following Ref~\cite{2016MNRAS.455L..72M}, we use
\begin{equation}
    \frac{{\rm d}^2 n}{{\rm d}z \,{\rm d}\! \log_{10} {\cal M}} =
    \dot{n}_0 \left[ \left( \frac{{\cal M}}{10^7 M_\odot}\right)^{-\alpha_{{\cal M}}} \mathrm{e}^{-{\cal M}/{\cal M}_\star} \right]  \left[ (1+z)^{\beta_z} \mathrm{e}^{-z/z_0} \right] 
    \frac{dt_{R}}{dz}\,,
    \label{eqn:model}
\end{equation}
where $t_{\rm R}$ is the time in the source frame~\cite{2001astro.ph..8028P} and we assume the Planck18 standard cosmology~\cite{2020A&A...641A...6P}.  
This model contains 5 parameters: $\alpha_{{\cal M}}$ and $\log_{10}{\cal M}_\star$ determine the shape of the mass function, $\beta_z$ and $z_0$ determine the redshift evolution, and $\log_{10} \dot{n}_0$ is the BBH merger rate density. Flat priors were used on these parameters in the ranges $(-3,3)$, $(6,10)$, $(-2,7)$, $(0.2,5)$, and $(-20,3)$ respectively.

First, we report the results of an analysis of the NANOGrav 12.5 year data using a BBH-only model spectrum. 
The 5 parameters of the BBH spectrum model were sampled over using the approach described in Methods section \ref{sec:results}.
The resulting posterior distribution revealed strong parameter degeneracies, consistent with the results in Ref~\cite{2021MNRAS.502L..99M}.
However, the total GW energy density of the background can be measured; the top panel of Fig.~\ref{fig:BBH_PT} shows the posterior obtained on the GW energy in the PTA frequency band assuming a BBH origin for the signal. The derived quantity $\Omega_{\mathrm{GW},\,\mathrm{BBH}}$ is computed as a function of the 5 parameters in the BBH spectrum model according to equation \ref{eq:Omega_gw} with frequency limits $f_{\rm low}^{-1}=12.5\,\mathrm{year}$ and $f_{\rm high}^{-1}=2.5\,\mathrm{year}$.

Secondly, we report the results of another analysis of the same NANOGrav data using a PT-only model spectrum. 
We use the PT model described in the main text that depends on the 4 parameters $\log_{10}H_\star/\beta$, $\log_{10}T_\star$, $\log_{10}\alpha_\star$, and $\log_{10}\eta$ for which we use the same priors as before. 
This is a repeat of the analysis described in Methods section \ref{sec:results}. 
Again, the total GW energy density of the background can be measured; the right panel of Fig.~\ref{fig:BBH_PT} shows the posterior obtained on the GW energy in the PTA frequency band assuming a PT origin for the signal. 
The derived parameter $\Omega_{\mathrm{GW},\,\mathrm{PT}}$ is computed as a function of the 4 parameters in the PT model for the spectrum model with the same frequency limits.

Finally, we report the results of a joint analysis of the data using a model for the spectrum that is a combination of both the BBH and PT models. 
This model depends on 9 parameters for which we use the same priors as before. 
This model is flexible enough to allow an arbitrary mixture of GWs from BBH and PT sources.
The 9-dimensional posterior on this model contains extremely strong parameter degeneracies hindering interpretation of the results. 
However, we can measure the GW energy densities in the PTA frequency band from the two components of the background; the red histogram in the main panel of Fig.~\ref{fig:BBH_PT} shows the 2-dimensional posterior on the GW energy densities $\Omega_{\mathrm{GW},\,\mathrm{BBH}}$ and $\Omega_{\mathrm{GW},\,\mathrm{PT}}$.
We also repeat this joint analysis with the inclusion of an integrated low-frequency constraint on the GW background, $\Omega_{\rm GW}<10^{-6}$, the results of which are shown with cyan contours in the main panel of Fig.~\ref{fig:BBH_PT}. 
The incorporation of the integrated constraint now rules out 10\% of the 9-dimensional posterior probability volume compared to the analysis using PTA data alone (the squared Hellinger distance between the two 9-dimensional posterior distributions is $d_{\rm H}^2 = 0.05$), although this small improvement is not clearly visible in Fig.~\ref{fig:BBH_PT}. These analyses highlight the difficulties of distinguishing different physical origins for the ultra-low frequency GW background.

\section*{Extended Data Section}\label{app:plots}

The extended data figure shows the same results as Fig.~\ref{fig:Om_plot} in the main text transformed into other commonly used quantities in the literature. 

\begin{figure}[h]
    \begin{center}\includegraphics[width=1\textwidth]{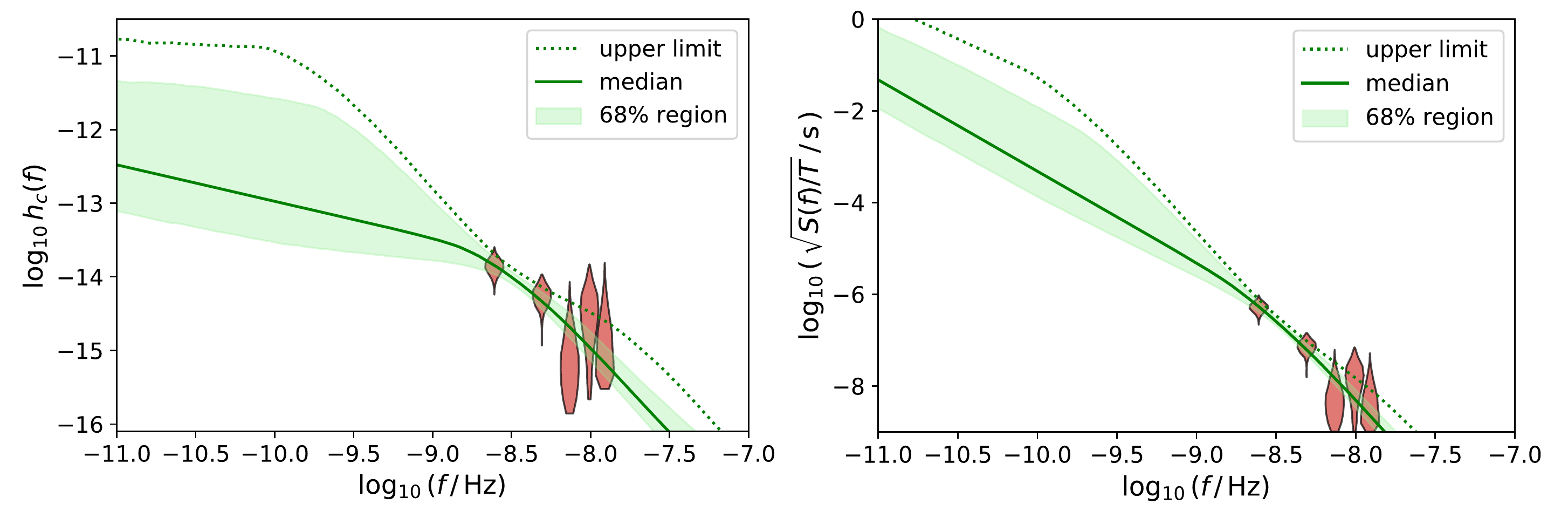}\end{center}
    \vspace{-0.3cm}
    \noindent \textbf{Extended data figure: PTA Constraints on the characteristic strain and PTA delay spectrum for a phase transition background.}
    Transformed posteriors on the PT spectrum shown in Fig.~\ref{fig:Om_plot} that were obtained from the 5 lowest frequency bins of the NANOGrav free-spectrum model (red).
    Line and shading style is identical to that in Fig.~\ref{fig:Om_plot}.
    Left: the characteristic strain, $h_{c}(f)$, defined in equation \ref{eq:omega_gw}. 
    Right: the delay, defined as $\sqrt{S(f)/T}$, where the power cross spectrum $S(f)$ is defined in equation \ref{eq:S_f} and $T=12.5\,\mathrm{years}$ is the duration of the observations.
\end{figure}


\bibliography{references}

\end{document}